\documentclass[amssymb,12pt,showpacs,aps]{revtex4}
\usepackage{graphicx}

\begin{document}
\title{Topological coding in hippocampus}
\date{\today}
\author{Yu. Dabaghian$^1$, A. G. Cohn$^2$, L. Frank$^1$}
\affiliation{$^1$Department of Physiology,
Keck Center for Integrative Neuroscience,\\
University of California,
San Francisco, California 94143-0444, USA, \\ 
e-mails: yura@phy.ucsf.edu, loren@phy.ucsf.edu}

\affiliation{$^2$ School of Computing, University of Leeds, UK; 
e-mail: a.g.cohn@leeds.ac.uk}


\begin{abstract}
The proposed analysis of the currently available experimental results
concerning the neural cell activity in the brain area known as hippocampus
suggests a particular mechanism of spatial information and memory processing. 
Below it is argued that the spatial information available through the analysis 
of the hippocampal cell activity is predominantly of topological nature. It 
is pointed out that a direct topological analysis can produce a topological 
invariant based classification of the cell activity patterns and a complete 
topological description of animal's current environment. It also provides a 
full first order logical system for local topological reasoning about spatial 
structure and animal's navigational strategies.
\end{abstract}
\pacs{87.10+e, 87.18-h, 87.19-j, 87.90+y}
\maketitle

\section{Introduction}
\label{section:intro}

The task of space perception and spatial orientation is one of the most
fundamental tasks faced by animals. The animal perceives itself and the
surrounding environment, plans and executes its movements and its behavior
in the context of the space that it experiences through neural activity in
its brain. Currently, there exists significant experimental data concerning
the mechanisms of spatial encoding based on electrophysiological recordings
from human, primate and rodent (notably rat's) brain. These experiments suggest
a number of specific principles according to which computation about space
in the brain is organized. Below we review some of this physiological and 
experimental data regarding the structure of neural space processing with 
the aim of introducing a certain view on the principles of space discernment 
in biological neural networks. This approach emphasizes that the phenomenon 
of space perception does not reduce to a passive reflection of the spatial 
organization of the external stimuli, but primarily is based on an active 
construction of the brain's own internal spatial framework.

The paper is organized as follows. Section \ref{section:properties} 
describes some basic experimental facts about neurophysiology of space coding 
and outlines some of the current paradigms used for analyzing neural activity 
patterns. A short supplemental list of physiological properties of the hippocampal 
neural cells is given in the Appendix. Section \ref{section:inner} discusses space 
description and space perception tasks in general terms, and outlines a specific 
structure of spatial information analysis that will be used in the rest of the 
paper. In Section \ref{section:partitions} we formulate a specific spatial 
information analysis task that motivates a particular approach to studying 
hippocampal space coding activity. It allows us to propose a hypothesis about 
the topological nature of the hippocampal space coding mechanism, which is 
further discussed in Section \ref{section:stability} in light of data obtained 
in continuously deforming environments. A discussion of the experimentally 
studied bounds of the continuous change regime is given in Section 
\ref{section:remappings}. 

An analysis of the PC population responses in Section \ref{section:coherence} is 
provided to support the claim that activity of these cells is globally coherent 
and may produce the emergent phenomenon of a physiological spatial frame (the
``inner space'') that serves as the basis of animal's spatial awareness. The 
topological properties of this space are analyzed in Section \ref{section:algtopology}, 
where it is shown that its invariant characteristics match the characteristics 
of the observed experimental environment. Section \ref{section:rcc} introduces 
the idea of using qualitative space representation analysis of the spatial 
information, in particular the Region Connection Calculus that allows the logic 
of animal's spatial behavior both in static and in slowly changing environments 
to be followed. A short discussion in Section \ref{section:discussion} puts this 
study into perspective of other theoretical analyses of neural space coding 
mechanisms.

\section{General Properties of the PFs}
\label{section:properties}

\begin{figure}[tbp]
\begin{center}
\includegraphics{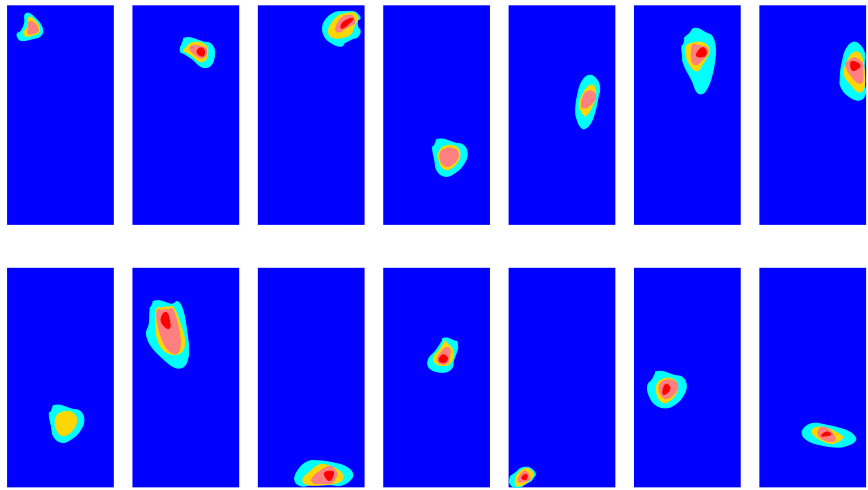}
\end{center}
\caption{A schematic representation of the place fields of 14 place cells 
in a square environment. Color code represents the increase of the firing 
rate from the background (blue) to highest level (red).}
\label{fields}
\end{figure}

In electrophysiological experiments, functional properties of the neurons and 
the neural networks are identified by studying statistical correlations between 
neural activity patterns in wake animals and various external (sensory) and 
behavioral parameters. One such functional property of neurons was found in 1971, 
when O'Keefe and Dostrovsky discovered that the firing activity of the pyramidal 
cells in a rat's hippocampus has clear spatial correlates \cite{OKeefe0}. 
Specifically, it turned out that these cells become active only in a relatively 
small portion of the environment and remain basically silent elsewhere (Fig.\ref{fields}). 
Hence these cells (called ``place cells'', PC) highlight a certain system of regions 
(called ``place fields'', PF), i.e. define a system of spatial ``tags'' via their 
firing activity.
The exact organizational and computational principles that produce place
specific firing patters or different PF partitions of the environment are
still not known, however there exists a general consensus about the overall
purpose of the PCs. It is thought that the PCs discretize continuous flow of
the sensory input into an inhomogeneous map, by using qualitative features of
the environment. The behavioral and functional significance of the hippocampus
has been demonstrated in variety of experiments. It has been shown that if the
hippocampus is partially or completely damaged, impaired or knocked out, the 
animal looses its full ability to solve many spatial navigation tasks, 
especially tasks based on following sequences of cues and retrieving
sequential (episodic) memories \cite{Eichenbaum1,Kesner1,Kesner2,Sharp}.
Experimental evidence indicates that the collection of the PFs completely
covers the whole environment that is recognized by the animal as ``familiar'',
and that the PFs reflect the structure of rat's internal map of this
environment. It is believed \cite{Sharp,Eichenbaum1} that this map serves as
one of the key structures of rat's spatial awareness, which was therefore named 
``cognitive map'' by O'Keefe and Dostrovsky \cite{OKeefe}.

\begin{figure}[tbp]
\begin{center}
\includegraphics{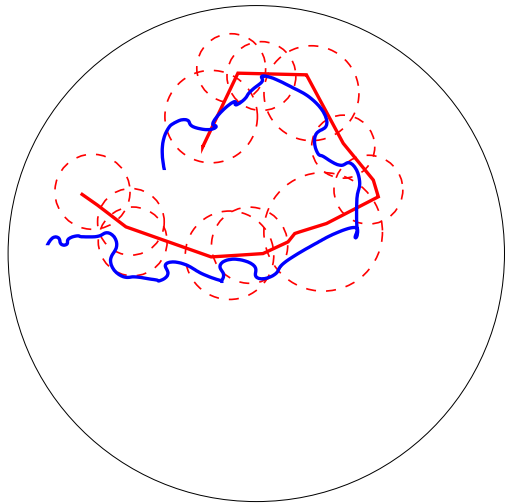}
\end{center}
\caption{Schematic representation of the trajectory reconstruction based 
on the place field activity \cite{Frank1}. Blue line represents the actual 
trajectory or the rat, red circles show the recorded place fields, the red
line represents the reconstructed trajectory. The accuracy of specifying rat's 
current position depends on the number of PFs with known locations in the arena 
and their sizes.}
\label{trajectory}
\end{figure}

It was also shown in \cite{Brown1,Wilson2} that knowing the positions of a
relatively few (70-80) PFs in a small (about 1 $m$ across) environment, one
can predict the rat's location at any time with an impressive accuracy
(Figure \ref{trajectory}) based on the current pattern of the activity of its PCs.
A simple extrapolation of the results of this experiment suggests that an
external observer should also be able not only to reconstruct the rat's
current location and trajectories as in \cite{Brown1,Wilson2,Lisman1}, but
also to characterize the whole space as it is encoded by the hippocampus, 
based on the information contained in the PC activity configuration. We 
hypothesize that the analysis of the PC activity should provide an insight 
into the organization  of the inner or ``physiological'' (in the terminology 
of \cite{OKeefe}) space that emerges from the spiking activity of the neurons.

Usually, the analysis and the interpretation of PC firing patterns is based
on correlating the firing events with the features (e.g. geometric) of the
physical location where firings have occurred. Thus, the properties of PFs
are usually defined in terms of the sizes and the shapes of areas where the
corresponding PCs are active. This perspective assumes an external observer.
From the point of view of the regions receiving PC input, proper description 
of the spatial representation of the environment should not depend on the 
external characterization of the firing patterns and rely only on the intrinsic
information encoded in the temporal structure of neural activity. From such
a perspective, understanding the mechanism of space perception depends
primarily on the possibility of interpreting the spiking activity from the
``proper'' point of view of the system, i.e. understanding the meaning of the
computation in the hippocampal network as it translates the {\em temporal}
pattern of the firing activity into a {\em spatial} pattern of the firing 
fields.

\section{Inner space and the spatial perception task}
\label{section:inner}

In this discussion, the term ``inner space'' is used to emphasize the point
that the spatial representation in the brain does not reduce to a passive
reflection of the sensory input. Rather, the state and the activity of the
brain bring about a separate {\em emergent} phenomenon of an inner space, 
that is an object in its own right. It is this emergent physiological inner 
space, rather than an ``objective physical external space'' that is directly 
experienced and perceived by the animal, in which the animal actually plans 
and executes its behavior, navigational tasks, etc.

There is a profound difference between the neural activity merely {\em 
reflecting} the external input by producing reliable responses, and an
activity pattern that {\em amounts} to having a separate internal 
representation of a space, defined in a certain specific mathematical sense. 
Clearly, the full task of space representation requires a variety of different
computations that integrate sensory information into several complementary
space representations. The analysis of the hippocampal PC activity outlined
below suggests that the PF map may provide the most basic, topological level
of the inner space coding.

The precise mechanism of space coding and even the scope of computational
tasks that the animal must address in order to achieve a sufficiently
complete space representation is unknown. However, from the mathematical
point of view, the colloquial term ``space'' implies a complex assembly of
several conceptually different structures, that must be coherently brought
together to produce the familiar realm of ``space''. These structures include:

\begin{enumerate}
\item Topological order, i.e. the relationships of adjacency and spatial
connectivity, spatial interior, boundary, closure, which allow the most
general and a highly abstract representation of the notion of spatial
continuity. Topological properties are preserved through deformations,
twisting and stretching of the space. They express whether the space can be
separated into parts, or contains ``holes'', etc., and are not sufficient 
to express ``sizes'' or ``shapes''.

\item Affine structure -- the possibility to define directions (vectors) at
a given point and a possibility to relate directions at different points in
a way consistent with the adjacency relationships and scale at different
locations in space. The ability to measure arcs, i.e. the angles between two
directions. The possibility to combine consistently the angular and the
distance measure. This is necessary to define geometrical objects, e.g.
straight lines, circles, and in general -- forms, contours, rigid bodies,
and the relationships between them.

\item Metric information -- a quantitative description of various spatial
scales via a topologically consistent distance measure between objects. This
construct allows the introductions notions of ``more close'' or ``less close'',
rather than just purely topological ``adjacent or not''. Establishing a global 
metric amounts to imposing a globally consistent system of scales at different 
points, i.e. the general notion of size.

\end{enumerate}

It is not clear a priori that this or similar stratification of the spatial 
coding task is implemented in the brain. However, it will be argued below 
that the task of encoding a topological representation of the spatial order 
is addressed by the hippocampal PCs, whereas other brain systems provide metric 
and affine information in order to provide the animal with a complete spatial
representation of the environment. The latter include the cells that signal
the instantaneous head direction of the animal regardless of the location of
the animal in the environment \cite{Muir,Taube1,Taube2,Ranck}. The
information about an animal's orientation, direction and duration of its travel
is also represented in the {\em ego}centric spatial frames encoded in the
parietal cortex \cite{Andersen1,Andersen2,Andersen3,Burgess3}. Metric
information is likely provided by proprioceptive (feedback from muscles and
joints) and idiothetic (self motion) cues \cite{Terrazas}, based on visual,
vestibular \cite{Mittelstaedt} sensory inputs. In contrast, the hippocampus, 
which is functionally the highest associative level network in the brain 
\cite{Lavenex,Banquet}, encodes and supplies the most abstract
representation of the space -- the allocentric topological map, that serves
as a ``locus'' \cite{Burgess1} of an animal's spatial awareness.

Importantly, there exist certain mathematical frameworks that allow 
the
question in what sense firing of the neurons in various parts of the brain 
may {\em amount} to the space that the animal perceives to be addressed. In 
the familiar approach, commonly used in mathematics and mathematical applications 
\cite{Alexandroff,Bourbaki,Kelly}, the space is understood as a certain proximity, 
scale and affine structure defined on a set of elementary ``locations'' -- the 
points of the space. In such an approach, the topological spatial structure 
emerges as a matter of associating subsets of points (originally unrelated 
spatially) into a consistent system $\mathfrak{U}$ of
``neighborhoods''. The consistency conditions (Kuratowski-Hausdorff axioms 
\cite{Alexandroff}) require that arbitrary unions and finite intersections
of the subsets from $\mathfrak{U}$ never produce subsets that lie outside of 
$\mathfrak{U}$. This guarantees that the chosen subsets can actually be
considered as ``proximity neighborhoods'' in the conventional geometrical
sense and that they generate a topological space structure on the original
set of points.

Interestingly, the proximity structure does not have to be applied to a set of 
infinitesimal elementary locations. Instead, the neighborhoods {\em themselves} 
can be understood the primary objects, so the proximity relationships are imposed 
on {\em regions}, rather than points. In such approaches (unified under the name 
``pointless topology'' \cite{Weil,Efremovic,Whitehead,Naimpally,Johnstone1}, or 
alternatively by the term ``mereotopology'' \cite{CohnVarzi03}) points are 
secondary abstractions, produced by intersections of a sufficient number of regions. 
The analysis of the topological connections between the regions in point-free spaces 
reveals a particular structure of logical and algebraic relationships between them 
that define the spatial organization. As a result, the pointfree topological space 
emerges from a logical/algebraic (``{\em locale}'' \cite{Johnstone1,Vickers}) 
structure imposed on the infinite set of regions 
\cite{Johnstone1,Johnstone2,Johnstone3,Laguna,Mormann,Roeper,Vickers,Whitehead}.

It is clear however, that the quasitopological inner space generated through
the activity of a finite number of the PCs in the hippocampal network can be
defined in terms of only a finite set of regions or spatial relationships.
In this sense, the inner space appears as a qualitative, finite approximation 
to the idealized continuum space. Physiologically, both the set of spatial 
regions and the continuity relation are derived through some discretized 
representation of the sensory input, and then used to internalize the external 
sensory structure in the form of a relatively coarse, qualitative space. In 
particular, the quasitopological aspect of the animal's inner space seems to 
be founded in certain qualitative region-based spatiotemporal reasoning schemes 
that provide discrete, qualitative versions of the standard set-theoretic or 
abstract ``pointless'' space computations.

Overall, the task of representing the space through neural activity can be
considered as an interesting and practical (also empirical) case of
mathematical constructivism, in which one aims to describe the emergence of
a topological space based on the firing mechanisms of neural cells. The
pointfree, region based space construction approach seems to apply directly
to the task of space coding in biological networks, which compute e.g. the
hippocampal quasitopological frame of the inner space from finite regions --
PFs.

\section{Place Field partitions}
\label{section:partitions}

As described in the previous section, the geometrical information contain in 
the observed system of PF can be used to reconstruct an animal's navigation 
paths and potentially describe certain aspects of the space as a whole based 
on the current PC activity. It is important however, that the information 
contained in the PF partitions alone is insufficient to represent the full 
extent of spatial relationships that are used and perceived by the animal. 
Some examples of spatial relationships that are {\em not} encoded via a ``place 
tag'' system are e.g. the observed sizes of the PFs themselves, the distances 
or the angles between PFs or between the external cues. In order to obtain 
such characteristics, the PC information must in general be complemented by 
additional information, e.g. rat's speed, the direction and the duration of 
straight runs and turns, etc.. In order to address the nature 
of the spatial information encoded in the PFs, it will convenient in the 
following discussion to refer to the following

\begin{itemize}
\item {\bf Space reconstructing thought experiment (SRE)}: Imagine a rat
that is running on a certain arena, separated from the experimenter (the
observer) in a locked room. The experimenter receives the real time signals
from the electrodes that are implanted into the rat's hippocampus, and is
free to analyze the recorded information in any way in order to extract from
it as much information as possible about the geometry of the arena and about
the navigational task faced by the rat.

\textbf{Question:} how much would the experimenter be able to deduce about the 
geometrical and spatial properties of the environment based on the PC activity?
\end{itemize}

{\em Comment:} The firing patterns of the PC can in principle contain more
information than provided by the spiking probabilities alone. Additional
navigational and spatial information, such as the size and the shape of the
PFs, etc., can potentially be supplied through other parameters, such as the
structure of the spike trains of single or multiple PCs, temporal structure 
of the cell bursting, or additional PF ensemble correlates. However, 
currently there is no evidence that there exist correlations between such 
parameters and the angle or the scale coding. With these stipulations, we 
will disregard below such ``fine structure'' coding of the spike trains.

With this understanding of the PF functionality, let us start by considering
the SRE task in the simple case of a rat running on a linear ($1D$) track.
After observing the PC signals for a sufficiently long time, the observer in
the thought experiment described above will notice that there is a linear
order to the time intervals in which hippocampal cells are active. Assuming
that the experimenter knows that the firings are spatially correlated (i.e.
that he in fact deals with the PCs), this would lead to the conclusion that
the environment is linear. In more complex ($2D$ or $3D$) environments the
correspondence to the spatial order is less direct, however a careful
analysis of the firing activity of a sufficient number of cells (for more 
formal discussion see Sections \ref{section:rcc} and \ref{section:algtopology}) 
should suggest to the observer that the temporal pattern of spikes is consistent 
with a possible ordering of regions in {\em a} space, i.e. that the firing events 
can serve as a consistent system of spatial location labels.

It is important to notice however, that given PF variability, this system of
spatial location tags a priori provides the observer only with {\em spatial
order} relationships and does not contain in itself any information about
the scale, the size or the shape of the environment. For example, in the
case of a SRE analysis of a $1D$ track it will be impossible to say, through
a mere observation of the PC activation sequences based on the properties of
the PFs listed above, whether the track is straight or bent (i.e. whether it
is I-shaped or U-shaped or C-shaped or S-shaped or J-shaped), what is the
scale of the environment, i.e. how long is the track or what is the
curvature scale of its sections. In order to produce a more complete spatial
description, the information about the sequence of locations should be
associated with the scale and the angle information. As mentioned above,
such information can be deduced from supplementing the time courses of the
PF firing with the information about the speed, direction and duration of
motion, etc., all of which do not correlate directly with the PF locations.
Based on the PC average activity profile alone, one can only determine
whether the track loops (O-shaped) or how many open ends it has (i.e.
whether it is X-shaped or J-shaped, however J-shaped and U-shaped tracks are
indistinguishable).

In the absence of any evidence of direct geometrical information contained
in PC temporal coding, we arrive at the hypothesis that the hippocampal
place cells encode predominantly the {\em topological arrangement of the
spatial locations} -- the topology of the inner ``physiological'' space.

It should also be mentioned, that although many of the shapes discussed above 
(e.g. W and I) are topologically identical, at the level of structural (e.g. 
sequential) organization of the available spatial locations, it is possible 
to separate each shape into connected parts, and to describe the topology of 
the assembly. The fact that the sum of some pairs of parts can be shrunk to a 
point but others cannot is the distinguishing feature, discussed in more detail 
in Section \ref{section:rcc}. As mentioned above (item \ref{partition} of the PF 
properties list in the Appendix), truck junctions and other distinguishing 
features in the space are typically explicitly coded for by the CA1 place cells 
\cite{Burgess2,OKeefe2}. In the context of the topological space coding, these 
features effectively play the role of spatial singularities and ``marked points'' 
that limit the ``topological plasticity'' of the space.

In view of this hypothesis, it is also significant that the hippocampus is known
to be largely responsible for ``sequence coding'' on a variety of different
time scales, even on the cognitive/behavioral level. Experimental evidence
indicates \cite{Agster,Bunsey,Chiba,Dusek,Eichenbaum,Fortin,Jensen,Kesner1,
Kesner2,Melamed,Schmitzer,Wallenstein}
that the behavioral performance of the animals with hippocampal lesions in
goal-directed sequence tasks is significantly reduced compared to the
control animals. Hence it appears that the hippocampus supplies current
representation of the consistent discrete sequential structure of the
environmental features, navigational cues and behaviorally relevant memories.

In a spatial context, such a regime of memory processing is in effect synonymous
to representing topology. It is well known that the topological structure of
a space can be defined not only descriptively, via a set of topological
invariants \cite{Novikov}, but also constructively, via an explicit, detailed 
set of discrete connectivity relationships between regions of the space 
\cite{CohnVarzi03,Efremovic,Johnstone1,Johnstone2,Johnstone3,Laguna,Mormann,
Naimpally,Randell,Roeper,Vickers,Weil,Whitehead}), which appears to be the 
basis of the hippocampal coding mechanism.

\section{Stability with respect to gradual changes}
\label{section:stability}

Interesting evidence in support for the topological coding hypothesis
comes from the analysis of the PF responses to the alterations of the
environment. Remarkably, if the external parameters, the ``features of the 
environment'' are changing sufficiently
slowly, gradually, then neither the number of PFs nor their relative order
change, only the exact location of PFs and the shapes of PC's activity
distribution profiles, i.e. the firing frequencies. So there exists a
certain regime of continuity in the representation following sufficiently
smooth changes in the external world. A number of studies have demonstrated
that gradual changes of local and distal cues in the familiar environment
produce a variety of orderly responses from the PFs, while maintaining their 
{\em relative} positions with respect to external cues and each other. For
example, a sufficiently slow dimming of lights, or gradual addition of
odors, or a smooth changes of the floor texture or a combination of such
changes does not produce changes in the mutual order of the PFs. If a rat is
taken from one square arena to another, similar enough for the rat to
recognize the similarity, then the PF mapping of the new environment remains
the same \cite{Lever}. In other words the operating regime of the network,
its synaptic strength configuration $S^{(i)}$, observed via PF layouts, 
does not undergo a major restructuring in order to track these changes.

It should be noticed however, that due to the fact that the environment 
transformations in a typical experiment involve many parameters that may 
affect miscellaneous aspects of hippocampal and cortical information 
processing, different functional responses in such experiments are often 
mixed together, which obscures regularities in PF behavior. For example, 
it was shown that moving separate objects in the arena, or creating a 
particular sensory cue dissociation \cite{Muller1,Fenton}, as well as a 
special animal training protocol \cite{Leutgeb} may lead to a variety of 
responses that sometimes violate the original spatial pattern or the 
firing regime of the PF (e.g. rate remapping, \cite{Hoz,Leutgeb,Leutgeb1}). 
However precisely because of the structural and functional heterogeneity 
of the external input, these variations usually do not directly support 
or oppose the topological coding hypothesis.

Since the hippocampus is known to be involved into complex memory formation
and consolidation processes in interaction with other brain parts, e.g. the
neocortex, it is clear that the implementation of each hippocampal PF map
may reflect various information about the environment, the objects within it
as well as arbitrarily complex relationships between them. Therefore, in the
context of studying specifically spatial aspects of hippocampal activity, 
it is important to specify a particular type of external transformation 
that can adequately address and bring forward specifically spatial aspect 
of the PC responses. For that purpose it is natural to consider spatially 
consistent, {\em geometric} changes, in which the whole set of sensory stimuli 
changes coherently, so that the mutual spatial order between every set of 
sensory cues is preserved and to study the of the PCs to this specific type 
of transformations. This approach is crucial for interpreting e.g. the 
phenomena of the partial (rate) remapping \cite{Leutgeb1} that appear as a 
result of separate cue manipulations.

The organized response of the PFs to the geometrical changes of the
environment was revealed in a particularly interesting group of experiments, 
\cite{Muller1,Fenton,Gothard1,Gothard2,OKeefe1}. In these experiments the 
shape of the environment is gradually altered (via a discrete sequence of 
transformations), and the positions of the PFs from the same PCs are 
recorded before and after the transformation. The results show that if 
the changes are sufficiently smooth, the positions of the PFs follow 
continuously the change of the geometry of the environment, i.e. they
drift continuously in the physical space without changing the original
relative activity structure. 

\begin{figure}[tbp]
\begin{center}
\includegraphics{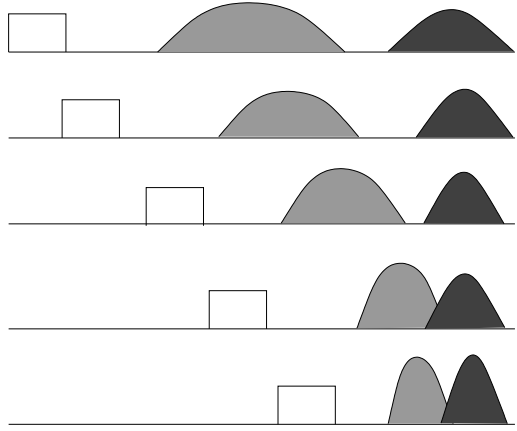}
\end{center}
\caption{Schematic representation of the elastic stretch of the PF layout
reported in \cite{Gothard2}.}
\label{gothard}
\end{figure}

More importantly, this connection is not local -- the responses of the 
place fields to the geometry alterations may be highly correlated across 
the whole environment. According to experiments of Gothard et. al 
\cite{Gothard1,Gothard2,Gothard3} the overall pattern of the PFs on a 
linear track shifts elastically in response to stretches or compressions 
due to the move of one of the track's ends (Fig. \ref{gothard}). 
Not only the PFs located next to the moving end respond by shifting, but 
also the rest of them, all along the track, shift accordingly to 
their distance to the moving end. The overall response pattern of PFs is 
as if they were drawn on an elastic sheet that can stretch or compress 
with the shifts of parts of the environment.

A particularly important aspect of these results is that the PFs are
not in fact anchored to a specific area of the environment. Instead, a
particular stable configuration of the hippocampal network seems to enforce
only the {\em relative} pattern of the PFs. The resulting global discrete 
map is in effect superimposed, projected onto the environment and that is 
preserved in continuous geometric transformations of the external cues. It 
is this hippocampal network configuration, and the corresponding structural 
pattern (a specific implementation of the cognitive map) that is directly 
available to the animal and may form a quasi-topological internal basis of 
an animal's spatial awareness.

This illustrates one of the core points of the proposed approach, namely that 
in a stable configuration of the hippocampal network, the activity pattern of
the ensemble of the PFs codes for the relative order, rather than the for 
association between the regime of elevated activity of a particular cell and 
a specific external location, object or an event.

Such a view allows to comment immediately on the property of allocentricity of 
the PFs, i.e. the independence of the PF locations on an animal's body position, 
behavior, etc. (see item (\ref{allocentric}) in the Appendix). Indeed, if for
a given cell, the location of its PF primarily reflects the {\em order} in which 
it fires with respect to the other cells in the network, the association between 
the regime of its elevated activity and a particular external region (its firing 
field) is a part of the global association of the whole hippocampal configuration 
with the structure of the environment. Clearly, if the relative activity regimes
of PCs in a given network configuration are fixed, the externally observed PF 
pattern does not change as long as the network configuration remains the same.
If an animal's behavior, its orientation and other similar characteristics do not 
affect the order relationships coded in the hippocampus, the activity of each 
cell is projected onto the same corresponding region, so the projected location
of the PF on the environment will be allocentric. In ``morphing'' environments, 
if the same configuration is projected onto a deformed arena, it creates the 
effect of the ``moving'' PF for an external observer.

If the geometric (or in some cases non-geometric \cite{Muller1,Muller2,Muller3}) 
transformation is a superposition of several dilations or homothetic 
transformations, there appear to be different groups of cells that follow a 
particular component of the deformation, thus forming a particular ``reference 
frame'' \cite{Gothard2}. In the experiment \cite{Gothard2} in which the walls 
of the environment were rotating while the floor and some objects scattered on 
it were static (or vice versa \cite{Muller1,Muller2}), one finds a subgroup of
cells whose PFs were bound to the walls (wall frame) and another subgroup of
PFs that remain floor \cite{Gothard2} or object \cite{Muller1,Muller2}
bound. For example, the PFs that are located close to the shifting wall
segments or moving barriers seem to be moving with them, so PFs are actually
tied to the elements of the environment.
\begin{figure}[tbp]
\begin{center}
\includegraphics{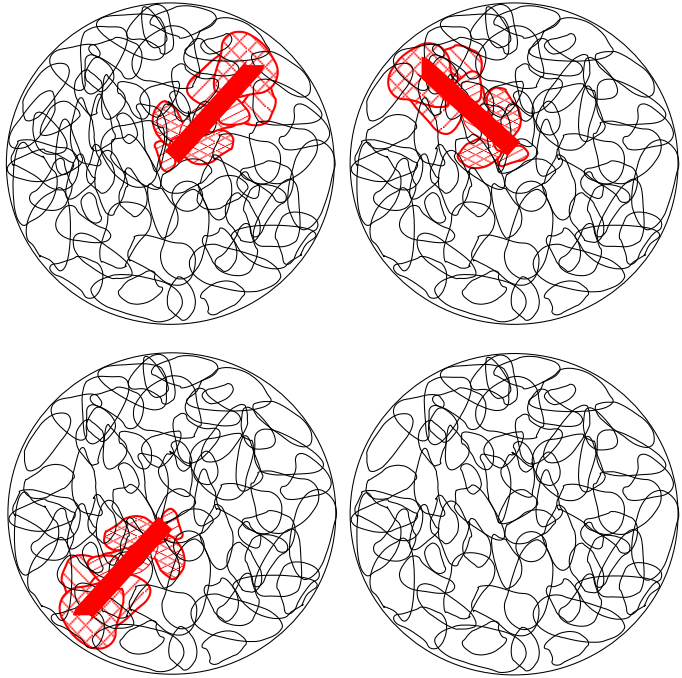}
\end{center}
\caption{Schematic representation of reference frames of place
fields in a circular arena with a moving barrier (red bar). Red 
shaded regions represent the barrier bound PFs that move with the 
barrier.}
\label{frames}
\end{figure}
This indicates that the space representation is layered in a certain way. 

Hence, based on the existing experimental evidence it seems reasonable 
to conjecture that in general the number of the frames is given by 
the number of the geometric components of the transformation and of 
spatial/behavioral modes. The latter possibility has been recently 
reported in \cite{Lee}, where it was observed that the closed linear
structure of the environment produced a separate frame -- a subpopulation 
of PFs that drift continuously along the closed path on a quasi-linear 
track without violating mutual order. The following discussion will focus 
either on geometrically simple transformations or on one such layer or 
a reference frame.

For the case of a single geometric transformation as in \cite{Gothard2}, it is
clear that since in the course of such changes the mutual order of the PFs
does not change, the spatial information that the PFs encode (from the point
of view of the SRE observer) remains the same. Therefore, although the
times at which neurons produced spikes may change, the SRE observer (the
hippocampal homunculus) can not follow the deformation of the environment
through this change. Hence, since the {\em structural pattern} of the
activity is preserved, i.e. the hippocampal network is encoding the same set 
of the relationships through sufficiently gradual geometric transformations 
and thus represents the same spatial information throughout the change. In 
other words, this type of the receptive field plasticity associated 
with the invariance of the mutual order of the receptive fields suggests that the
space representation by the hippocampal network is invariant with respect to
sufficiently smooth geometrical distortions of the environment, and hence
provides a certain {\em flexible} (for the external observer) discretized
representation of the behaviorally relevant spatial information. So in that
sense the PF patterns encode the ``elastic skeleton'' of the space represented 
in rat's brain, i.e. provide the information of topological nature.

We further hypothesize that results similar to Gothard et. al. will be found
for more general alterations: if the environment changes its geometric
configuration (stretches or bends) sufficiently slowly, then the PFs will
follow this change in the sense of \cite{Gothard2,Gothard3}, so the temporal 
{\em order} of the firing sequences will remain invariant.

The above arguments require certain additional stipulations about neglecting
the ``scale'' effects that may arise due to the finite size and the firing
profile of the PCs. For example, PFs have a certain range of characteristic
sizes, which imposes a restriction on the scale of the changes that may be
ignored by the hippocampal network: the scale of changes per cell should be 
less than or comparable to the characteristic size of the individual place 
field. Also, the excitation level of a PC increases as the rat is getting 
closer to the center of its PF, which may provide some {\em local} metric
and directional information, within the scale of (a priori unknown) size 
of the PF, which does not change the argument about the topological nature 
of the space coding in PF population. 

Overall, these results suggest that PFs code for the topological information 
in the conventional mathematical sense in at least in two aspects -- as a 
consistent collection of places that expresses spatial order, which (as a 
matter of empirical coincidence) has the ``elastic grid'' properties, i.e. 
is invariant to a certain range of geometrical transformations. Below the 
spatial discretization mapping will be referred to as {\em quasitopological}, 
to emphasize that the topological regime is stable only within a certain
range of parameters, which limits the scale and the time course of the
external changes.

\section{Remappings}
\label{section:remappings}

\begin{figure}[tbp]
\begin{center}
\includegraphics{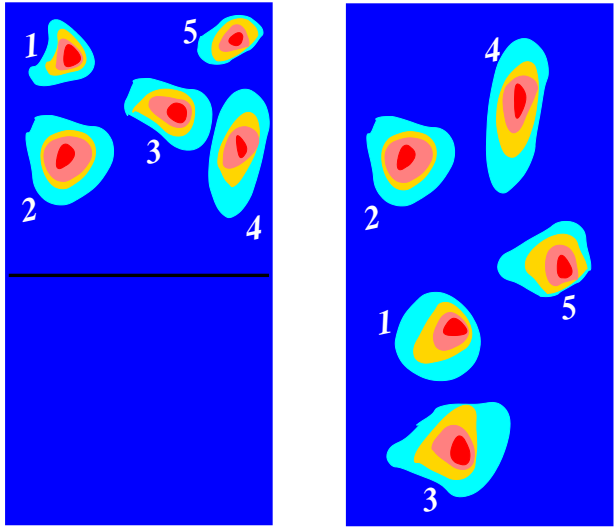}
\end{center}
\caption{Remapping caused by the global abrupt change of the environment
(Mumby box). The order of the 5 PFs in the upper half arena changes completely 
after the barrier (black line) is removed.}
\label{mumby}
\end{figure}

Experimental evidence also suggests that significant and/or rapid changes
may cause a complete change of the representation -- the so-called remappings, 
in which the old pattern of PFs is replaced by an entirely new one. In the
remapping process cells may completely change their firing properties: some 
previously quiet cells may become active, previously active cells may shut off, 
and the overall PF location pattern acquires a completely different structure. 
For example, in the experiments with a thin barrier placed into the arena 
(Fig.\ref{mumby}), the PFs on which this barrier was placed disappeared, i.e. 
the corresponding PCs stopped firing there, although the rat could physically 
visit most of the place occupied by the disappeared PF from both sides of the 
barrier. If large (compared to the size of the arena) barriers are added or 
removed, so that the geometry of the environment, e.g. the available navigational 
routes, change significantly (Fig.~\ref{mumby}), then the PFs pattern can remap 
not only in the vicinity of the barrier, but also in the rest of the arena 
\cite{Sharp,Save1,Save2,Fuhs2}. 
During such remapping processes, the strengths of the synapses may reset 
\cite{Muller3} and so the hippocampal network transitions into a different 
state $S^{(j)}$. Experimental evidence also indicates that abrupt geometric 
deformation can also cause remappings; however the
slower are the changes the larger is range of changes that do not cause a
major remapping.

The organized response to the external changes, in which the
quasi-topological order of the PFs is preserved and the abrupt scrambling of
the PF order represent two qualitatively different regimes of behavior and
of spatial coding. A simple example of remapping is provided by the
violation of the linear order of PFs on a $1D$ track following the abrupt
changes of the track configuration. The remapping here
corresponds to a violation of the linear order of the PFs, in contrast with
the orderly stretch observed in \cite{Gothard2}.

To distinguish formally the orderly shift of PFs from shuffling in more
general environments, a PF configuration can be associated with an
auxiliary graph $G$, defined e.g. as the Delaunay graph of the Voronoi
tesselations of the PF layout \cite{Edelsbrunner,Okabe,Tenenbaum}. 

For a given PC/PF, its Voronoi region is defined as the set of points on the arena
that are at least as close to the point of its maximal firing rate as to the
maximal rate point of any other PC. The Voronoi region of a given PC has no a
priori geometric relationship to its PF -- e.g. it may contain its PF or be
contained in it. Once the Voronoi diagram for a given PF layout is obtained,
its dual object, the Delaunay graph, $G$, is built by connecting the maximal
firing rate locations of neighboring Voronoi regions by an edge. With these
objects at hand, the regime of regular responses to the slow changes can
be defined when the graph $G$ is ``continuously deformed'', i.e. although the 
Voronoi regions change, the connectivity of $G$ does not, whereas remapping
corresponds to the case when the {\em topology} (connectivity) of the
graph $G$ is altered, i.e. the remapping can be defined as a transition
between two equivalence classes of the corresponding Delaunay graphs. 

This definition not only allows different partial descriptions of the
PFs structures that are due to the same state of the network to be
related to each other, but also to relate PF partitions that are
generated in different, but ``close'' states of the network. 

It is also clear that the correctly defined order should reflect the 
actual structure of the space rather than the structure of a given set 
of PFs that happened to be detected in a particular electrode configuration. 
This implies for example that if additional PFs are detected (e.g. by adding 
an extra electrode), the original graph $G$ should be considered as an
approximation to the extended graph $G'$, that is obtained by an
appropriate insertion of vertexes (new PFs) and edges \cite
{Bern,Edelsbrunner}. Hence the {\em spatial order} of PFs is defined as set
of Delaunay graphs defined up to a continuous deformation and ordered by
inclusion (graph extension).

This definition allows to relate not only different partial descriptions of
the PFs structures that are due to the same state of the network, but also
to associate some PF partitions that are generated in different states of
the network. It is clear for example that due to a finite size range of the
PFs in the experiment \cite{Gothard1}, the ordered chain of PFs, e.g. $PF_{1}$, 
$PF_{2}$, \ldots , $PF_{N}$ cannot keep up stretching indefinitely with the
space expansion. One would expect that at some point new PFs must appear in
order to cover the additional space. In another case, if some part of the $
1D $ track is inflated, then at some point the original linear order of PFs
will be substituted by a more complex $2D$ order. In the case of the
environment shown in Figure \ref{fields}, if another rounded area is gradually 
added somewhere in the middle, then 
one would expect that the original PF pattern within the growing section will 
initially inflate, but then (e.g. when the size of the added space will be 
comparable to the size of a typical PF), some new PFs will have to fill in 
the new place. This may happen with or without the overall order violation. 
If the global order on the originally available part of the track is preserved, 
this geometrical deformation in the middle will increase the number of PFs
and turn the linear order of the PFs on the straight section into a $2D$ 
order from the same equivalence class. Similarly, some PCs may shut off if 
their PFs are squeezed out of a contracting track, which also does not 
violate the spatial order. In either case, if the original pattern is only 
amended, not scrambled, by the new fields, the original topological order 
is preserved. According to the above definition such restructuring will not 
qualify as a remapping.

It should be emphasized that this definition of the PF order via graph
topology helps only to separate the regime of continuous changes from the
remapping events, however it does not necessarily specify either the state
of the network or mark the topology of the inner space. The graphs from
different equivalence classes may represent an inner space of same topology,
and the PF partitions generated by different states of the network can be
described by equivalent graphs. The problem of defining the topology of the
inner space can be addressed through analysis of the connections between the
PFs, which would require an entirely different analysis, outlined in the
Section \ref{section:algtopology}.

\section{PF coherence}
\label{section:coherence}

The properties of the PFs indicate that an external observer can build a
qualitative {\em quasi-}topological description of the space that reflects 
the internal structure of the space perceived by the animal. 

However, the statement that the rat itself possesses an explicit map of space
implies more than a mere availability of the information the locations or than
completeness of this information for an external observer. Having a set of 
space tags does not necessarily imply having an internal space, in the same way 
as having a database of postal addresses does not amount to having a spatial map 
that facilitates or even permits navigation and orientation. The complete space 
information provides a vast framework that allows a countless variety of nontrivial 
navigational tasks to be addressed, as well as the building of spatial consistency 
schemes,spatial planning, etc.

The information contained in separate PFs may not necessarily be organized
to produce an ``inner space'' \cite{Mackintosh,Redish2}. An alternative to 
possessing an emergent internal inner space produced by a specially 
coherent organization of the network computations might be a certain organized 
database of ordered links, connections, that are not necessarily geometrized
or geometrizable (analyzing the database of links between French internet 
websites will not lead to reconstructing the geographic map of France). The 
emergence of an inner space would imply that the system actually knows how to 
associate the PF regions together on a global scale to produce a self consistent, 
globally coherent, explicit spatial map. The main argument in support of 
quasitopological inner space in the hippocampus comes from studying the 
coherence of the PF collective responses to external changes, which show 
that the behavior of the ensemble of the PFs can be better understood on 
the level of the properties of the inner space as a whole, rather than an 
assembly of individual links between location tags. Hence the nature and 
the mechanism of connectivity between PCs is of great importance.

Previously mentioned facts about the collective and coherent responses of
the PFs to the changes of the environment, either in a local spatial frame
or especially in global shifts or moves of the PFs as in \cite{Gothard2} as
well as global remappings in response to relatively local changes, strongly 
suggest that PC activity should always be considered in the context of the 
global state of the network and collective behavior of the PFs. Interesting 
evidence is provided by the observation of replays of the correct (direct or 
reversed) sequences of PCs in wake \cite{Foster} and asleep \cite{Louie,Skaggs,
Wilson1} animals, which reflect the pattern of the same cells during actual 
spatial navigation. This result shows that correct PF sequences can be pinged 
by the system which remains in a state mapping state during animal's sleep.

An important support for functional connectivity between PCs comes from the
phenomenon of prospective coding, found in \cite{Frank2}. In this experiment
a rat was running on the ``W-track'' alternating the left and the right turns 
on its outbound journeys from the middle to the side arms.
\begin{figure}[tbp]
\begin{center}
\includegraphics{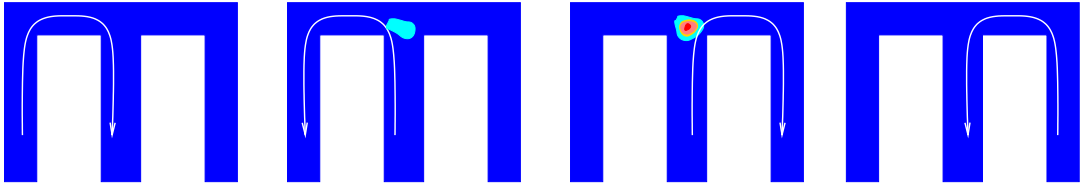}
\end{center}
\caption{A schematic representation of the activity of a CA1 place cell showing 
prospective coding on a W track \cite{Frank1}.}
\label{prospective}
\end{figure}
Interestingly, a cell with the PF located at the spot where the left and the 
right trajectories split, was active only if the rat was going to turn to the 
right, and the same cell would be silent at the same spot, if the rat's 
intention was to turn to the left. Such behavior indicates that PCs are not 
just ``place'' cells, but rather cells that code for a place in a particular 
spatial and sequential context, so some larger scale connections are involved 
in these cells' activity \cite{Battaglia,Ferbinteanu,Mehta,Muller4}.

Since deformations of the linear tracks do not cause remapping of the
PFs, it is likely that the cells exhibiting prospective firing in the arena
of particular geometrical configurations (e.g. W-track) will retain their
functional properties if the environment is gradually deformed. From the
point of view of the SRE observer such prospective firing does not provide
any additional information about the geometry of the environment. Instead,
it may indicate the existence of a functional connectivity between the PCs 
and unite continuously the PF pattern into a single space coding ensemble 
both in static configuration and through the gradual geometrical changes of 
the environment.

An additional mechanism for local and global synchronization of the PC activity 
is provided by the global EEG potential oscillations, such as the $\theta$ rhythm. 
A number of studies have indicated that the $theta$ rhythm helps organize sequential 
encoding and read off hippocampal information \cite{Hasselmo1,Hasselmo2,Lisman,Lisman1}.
The overall evidence suggests that one should regard the fact that a cell
is active in a certain region not just as mere ``place markings'', but rather 
as ``marking the location within a particular structured pattern of other
locations''.

At the cognitive level, structural coherence of the hippocampal memories is
manifested in the phenomenon of episodic memory, i.e. the ability to put a
specific memory into the context of preceding and succeeding events, as well
as the ability to produce complete memory sequences from a single structured 
input \cite{Eichenbaum}. It is well known that in humans damaged or lesioned 
hippocampus leads to severe impairment of these capacities, which implies that 
the hippocampus naturally embeds memory elements (e.g. memories of spatial 
locations) into a globally structured context. The pattern of difficulties 
faced by the animals with hippocampal damage in solving various cognitive 
tasks in changing environments shows that the activity of the hippocampal 
network is also essential for the ability to recognize persistent patterns. 
Even if the overall cognitive structure of the task and the general relational 
(topological) structure between salient features of the changing environment 
remains the same, animals (e.g. rats and humans) loose the effective ability 
to navigate among familiar cue patterns and to put separate cues and memories 
into the previously familiar general context.

The existence of a global map is also indicated by the phenomenon of the 
so-called ``path integration'' \cite{Etienne}, i.e. the ability of the animals 
to find a direct path back to the origin of their journey after traveling along
geometrically complex trajectories. However, path integration does not
appear to be a hippocampal function \cite{Alyan}, as the ability of path
integration is not impaired in animals with hippocampal damage, and reflects
a more general cognitive brain functioning \cite{Redish1}.

\section{Topological analysis of space coding}
\label{section:algtopology}

The above discussion of the nature of the PC space coding allows the space
reconstruction task in the SRE experiment to be specified by reducing it
to the task of extracting the quasitopological spatial frame encoded in the
spiking data.

Conceptually, the task of reconstructing the space {\em as perceived by the
rat} is different from the task of reconstructing the space that the SRE
observer expects to see. In the latter case the observer verifies the
consistency of the received rat's PC signals with his own representation of
the environment, and in the former case the goal is to establish the
internal consistency of spatial relationships assuming a specific space
coding mechanism. The hippocampal topological basis of the perceived space
can be understood as the state of the hippocampal network, that is directly
experienced by the animal and that is projected onto the observed
environment via PC activity as a partition of PFs. To emphasize this
distinction between the observed and ``inner'' spaces, the first task will be
referred to below as the {\em inner} space reconstruction experiment (ISRE).

In either case, the underlying assumption is that the firing fields are
associated with the neighborhoods $U_{i}$ of a topological space $X$ and
hence the PC information must be consistent with possible spatial
relationships between the neighborhoods of $X$. In the SRE context the
topological space in question is the environment $S$ as verified directly by
the observer, and in the ISRE case it is the topological skeleton of the
hypothetical ``inner space'' $\tilde{S}$ in which the animal perceives itself,
so the ISRE observer in effect induces the structure of the
``inner PFs'' covering the inner space of
the animal from the activity of PCs.

In both cases the analysis of the encoded space is based on the fact that
the topological properties of a space can be deduced from the properties of
its coverings by open sets $U_{i}$ (neighborhoods) defined either directly
or via some additional structures, associated consistently with the
neighborhoods. Such analysis allows the topology of the space to be
characterized in terms of topological invariants \cite{Novikov}. In the case
studied here, the structures associated with neighborhoods are the firing
rates of cell populations, which can be used to identify the topology of the
environment both in the SRE and in the ISRE contexts for different PF
layouts, and to argue that the topology observed by the experimenter is
equivalent to the topology of the rat's own inner space.

Mathematically, the task of establishing a global arrangement of locally
defined structures over a topological space $X$ is a well defined problem
that in its most general form is addressed in the so-called sheaf theory 
\cite{Godement,Hirzebruch,MacLane,Swan,Tennison}. The concept of a sheaf
captures the idea of associating the local information with the spatial
structure of a topological space as a whole \footnote{According to \cite{Swan}, 
``Sheaf is effectively a system of local coefficients over a space $X$.''}, 
and can provide a general framework for analyzing different types of neural 
information associated with the topological structure of the environment, 
such as the angular orientation at different locations (spiking rates of 
head direction cells) \cite{Taube1,Taube2,Sargolini} or the state of running 
straight or turning (cells in the parietal cortex) \cite{Burgess3,Nitz}. 
So from the perspective of the topology reconstructing task, the (I)SRE 
goal is to define the structure of the topological space based on the PC 
firing information, i.e. to associate the local PC information into a 
global topological characteristics of the ``inner space'' as a whole.

In the context of studying a rat's mechanisms of space coding, the PFs appear
to be primitive regions, in terms of which more complex spatial objects are
constructed, both smaller or larger than a single PF. Although it may be
possible that for the rat itself, spatial awareness is based on the analysis
of currently coactive PCs, the analysis made by the (I)SRE observer can
include any relationships between PCs/PFs that were observed in the course
of a (I)SRE experiment.

\begin{figure}[tbp]
\begin{center}
\includegraphics{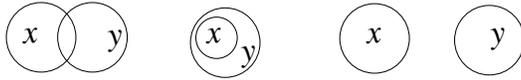}
\end{center}
\caption{Intersection of two regions $x$ and $y$, inclusion of regions and
separated regions.}
\label{cxy_rels}
\end{figure}

Let us assume that for a given set of $k$ cells and the corresponding threshold 
levels $\theta_{i}$, $i=1$, ..., $k\,$, the PF data is sufficiently complete, 
so that the PFs produce a complete covering of the environment,
\begin{equation}
\cup_{i}PF_{i}^{\left(\theta_{i}\right)}=E.
\end{equation}
The regions can be defined as the rat's position states in which the firing
frequencies, $f_{1}(t)$, ..., $f_{k}(t)$, of a selected set of PCs, $c_{1}$, 
..., $c_{k}$ vary within chosen limits, $f_{i}\in \Theta_{i}\equiv \left[ 
\theta_{i}^{\min},\theta_{i}^{\max}\right] $. The union, the inclusion and 
the intersection of the regions (Fig.\ref{cxy_rels}) can be defined immediately 
via the corresponding union, inclusion and intersection of the firing rate 
intervals $\Theta_{i}$ of two cell populations. 
For example, $x$ is a subregion of $y$, $x \subseteq y$, if whenever every 
place cell defining $x$ is firing at greater rate than its threshold, then
so is every cell defining $y$.

Since for example a region $V=PF_{1}\cap PF_{2}$ is contained in $U=PF_{1}$,
the restriction operator $i_{UV}$ from $U$ to $V$ corresponds to selecting the 
rat position states in which both cells simultaneously have a high firing rate.
Clearly, for the three nested neighborhoods $W\subset V\subset U$, the
restriction from $U$ to $V$ and then to $W$ (region of coactivity of 3 PCs)
and the direct restriction from $U$ to $W$, will yield the same result, i.e. 
$i_{V,W}\circ i_{UV}=i_{U,W}$. It is important that such analysis of the
``covering'' can be made directly in terms of the firing properties of the PCs
that does not refer to the externally observed geometrical PF features, and
does not require observing the behavior of the rat or knowing its
environment. It is also important that the restriction operation does not
depend on the restriction sequence. As a result, the firing rate of the
whole PC population can be extended in a unique way to form a continuous
function $f:U\rightarrow R$ which agrees with all the given $f_{i}$. This
provides the topological characterization of the space in terms of well
defined topological invariants, based on the analysis of its coverings.

With a given set of the regions $U_{1}$, ..., $U_{n}$, covering the environment, 
one can associate the following multidimensional polytope (simplex) also called 
the ``nerve'' $N$ of the covering \footnote{This term is purely mathematical and
has nothing to do with the uses of the word ``nerve'' in physiology and neuroscience}: 
1) The vertexes $\sigma_{i}^{(0)}$ of the simplex correspond to the individual regions 
$U_{i}$. 2) The edges $\sigma_{i_{1}i_{2}}^{(1)}$ correspond to overlapping regions, 
$U_{i_{1}}\cap U_{i_{2}}\neq \emptyset $, i.e. to the coactive cell populations. 3) 
The two dimensional simplexes (facets) $\sigma_{i_{1}i_{2}i_{3}}^{(2)}$ correspond 
to triple intersections $U_{i_{1}}\cap U_{i_{2}}\cap U_{i_{3}}\neq \emptyset$. 4) In
general, $k$--simplexes $\sigma_{i_{1}i_{2}...i_{k+1}}^{(k)}$ correspond to nonempty 
intersections $U_{i_{1}}\cap U_{i_{2}}\cap ...\cap U_{i_{k+1}}\neq \emptyset $, or a 
set of $k+1$ coactive PC sets.

In particular, one can consider coverings generated by the PFs themselves and the 
corresponding simplex generated by the peaks of activity of PCs, the overlapping PFs, 
$PF_{c_{1}}\cap PF_{c_{2}}\neq \emptyset $, the coactive triples of PCs 
$PF_{c_{1}}\cap PF_{c_{2}}\cap PF_{c_{3}}\neq \emptyset $, etc. 
\begin{figure}[tbp]
\begin{center}
\includegraphics{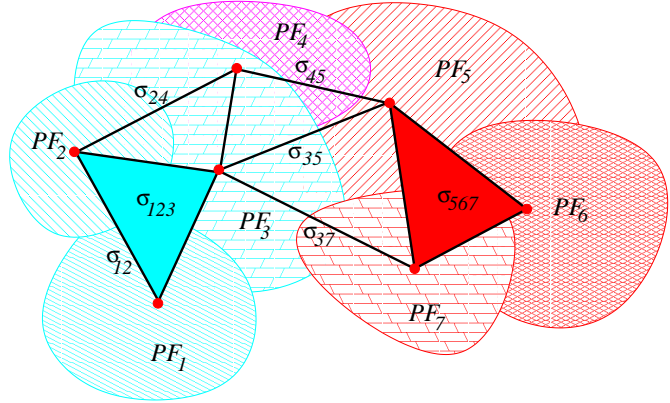}
\end{center}
\caption{Simplex generated by a covering of a plane environment.}
\label{simplex}
\end{figure}
The analysis of the (I)SRE observer is based on studying the topological
structure of the resulting simplicial complex using an appropriate system of
algebraic indexing of the simplexes. The topological properties of the space
can be revealed via the structure of the space of linear functions on the
simplexes, $\alpha ^{k}(\sigma_{i_{1}...i_{k+1}}^{k})$ -- the cochains
taking values in properly selected set of coefficients $F$, e.g. $0$ and 
$1$. The region restriction operation $i_{UV}$ allows the functions defined 
on $k$-simplexes to be put ``into the context'' of the higher dimensional 
$\left(k+1\right) $-simplex that they enclose, via the so-called ``codifferential''
operator $\delta $ \cite{Novikov,Swan,Hirzebruch}. The resulting algebraic
structure defines the set of topological invariants that uniquely
characterize the topology of the simplex -- the \v{C}ech cohomologies of the
covering, $H^{q}\left( N\left\{ U_{i}\right\} \right) $.

If additional PFs are observed in (I)SRE, the old nerve of the covering
will be inscribed into the new one, $\varphi_{UV}:N\left\{ U_{q}\right\}
\rightarrow N\left\{ V_{p}\right\} $, so the refinement of the simplex
induces a mapping of the topological indexes
\begin{equation}
\varphi_{UV}^{\ast}:H^{q}\left( N\left\{ U_{i}\right\} ,F\right)
\rightarrow H^{q}\left( N\left\{ V_{i}\right\},F\right).
\end{equation}
Such analysis allows the topology of the environment to be characterized
using nonequivalent (in the sense of the Section \ref{section:remappings}) 
coverings that are generated during the remappings. If both new and old 
coverings are defined via singly connected, contractible in the topology of
$X$ PFs, they define the nerve of the simplex of the same cohomological type, 
so different PF coverings of the same environment must provide equivalent 
topological representations of it. This also shows that unlike the topological 
order of the directly observed PF partition (i.e. the topological class of the
Delaunay graph) the topology of the inner space is not violated in remappings.

Furthermore, a basic (but important) result from the algebraic topology
states that in case if $X$ is a manifold and if the covering is such that
all $U_{i_{1}}\cap ...\cap U_{i_{k+1}}$'s are contractible, e.g. small
convex regions in the metric of $X$, then the nerve of the covering is
homologically equivalent to $X$,
\begin{equation}
H^{\ast}\left(N\left\{U_{i}\right\} \right) =H^{\ast}(X),
\end{equation}
so the topological (cohomological) characteristics of the covering are
identical to the topological characteristics of the manifold itself. This
simple result allows the (I)SRE observer to make judgments about the
topological properties of the environment as a whole and in particular to
conclude that the topology of rat's ``inner space'' coincides with the
topology of the environment defined via explicitly observed PF coverings.
This equivalence also shows that the hippocampal space representations of
different rats can be mapped one onto another.

It is interesting to mention that the results and the methods obtained above 
might be used for extracting the global spatial topological characteristics of
PC activity patterns not only from awake animals, but also from the PC replay 
data collected during animal's sleep \cite{Karlsson}.

Certainly, the possibility to compute correct topological invariants depends 
crucially on the quality of the covering, i.e. on the availability of a 
sufficient number of the PFs. Usually the environments in which the behavior 
of a rat is studied have geometrically simple form, so it is assumed for 
simplicity that the environment is covered by a sufficient number of PFs.
In case if the geometry of the environment is complex, the derivation of 
the correct cohomological characteristics becomes a more subtle problem \cite
{Benedikt,Carlsson,Delfinado,Dey,Gonzalez,Munkres,Robins1,Robins2,Zomorodian},
that can be helped by the local analysis of the spatial relationships encoded
by the PFs.

\section{Qualitative Space Representation. RCC}
\label{section:rcc}

The mathematical formalism used above for describing the {\em global} 
topological properties of the space does not however provide a framework 
for a practical, biologically plausible analysis of the current, local PC 
firing configurations. It does not necessarily suggest a scheme for a 
{\em local} spatial information processing that can serve as the basis 
of the local spatial planning, navigation and in general for {\em spatial 
reasoning}. On the other hand, there exists a variety of qualitative space 
representation (QSR) methods \cite{Cohn2}, which provide practically useful 
and conceptually complete approximate representations of space that include 
local topological information analysis. The reasoning techniques within 
different QSRs provide case specific formal languages and logical systems 
that include the notion of spatial proximity and other necessary spatial 
relationships. These languages can provide a complete description of arbitrarily 
complex spatial reasoning schemes used for navigating, spatial planning and 
establishing spatial (in)consistencies, which have been studied in a number 
of mathematical \cite{Randell,Cohn1,Cohn2} as well as applied contexts, such 
as organizing geographical databases, robotics, artificial intelligence, 
object recognition, etc. (e.g. \cite{Egenhofer92a,Escrig98,Fernyhough2000}) 
and including some bio-applications \cite{fois01,Donnelly06} but not to our 
knowledge, the task under consideration here.

A simple example of qualitative spatial reasoning is provided by the above
analysis of a linear sequence of PFs on a linear track, that allows the SRE
observer to conclude that the space the rat has explored is linear and how
many ends it has. If the cell population activity patterns are always
consistent with a simple linear (direct or reversed) sequence $a_{1}$, 
$a_{2}$, ..., $a_{n}$, then the SRE observer can conclude that the track is
linear. The violations of the linear order via $a_{1}\rightarrow a_{n}$ or 
$a_{n}\rightarrow a_{1}$ transitions only, signify that the track is circular,
while presence of ``forking'' sequences signals branching arms of the track, 
etc. (assuming no inconsistencies would be expected in actual biological 
recordings in stable hippocampal state). 

A more general form 1D reasoning is addressed by the so-called Allen Interval 
Calculus relations reasoning calculus \cite{Allen}. The trajectory 
reconstruction mentioned in the Section \ref{section:properties} also in 
effect represents a simple example of qualitative spatial analysis.

In general, QSR methods can be applied to representations derived from
analogical representations based on explicit maps or on database
representations of spatial relationships (e.g. statements like ``the book is
in the bedroom on the table that is to the left of the window''), as well as
on a combination of both. The advantage of the explicit map is that it can
provide (typically at a higher computational cost) a basis for generating
task specific spatial relations, while in the second (reactive) case the
relationships are not defined at all unless stated explicitly (e.g. if no
relationship is specified about where the table or the book are in relation
to the chair, there is no way to deduce that -- unless it can be inferred 
indirectly via the relationships to one or more other objects). Different 
QSR schemes could be used to represent different aspects of space coding in 
various brain parts \cite{Burgess3}. In the cognitive space coding schemes,
QSR methods may be used for describing the relationships between behaviorally 
relevant or otherwise salient spatial locations \cite{Corbetta,Duhamel}, i.e. 
in establishing the cognitive ``spatial salience map'' of the environment. 
On the other hand, the hippocampus seems to possess an explicit spatial map 
that produces the topological skeleton of the inner space.

A practically convenient QSR method for describing topological spaces is
provided by a pointfree method called the Region Connection Calculus (RCC) 
\cite{Randell,Cohn1,Cohn2,Gotts}, that can be used for studying the
quasitopological space that emerges due the hippocampal PC activity. In its
general form, the RCC method unifies several QSR schemes that are based on a
single primitive, binary, reflexive and symmetric connectivity relation 
${\sf C}(x,y)$ (region $x$ connects with region $y$), that relate every two 
regions of space. In the context of the (I)SRE 
analysis, ${\sf C}(x,y)$ can be evaluated directly as the coactivity of the 
cell populations above the corresponding threshold levels, hence the 
RCC analysis can be used directly in the space reconstruction setup.

From the point of view of the (I)SRE analysis, it is important that (see 
e.g. \cite{Cohn1,Cohn2,Gotts,Randell} and the references therein) that the 
RCC relationships form predicates of a {\em first-order logical system},
which turns the spatial reasoning based on the PF information into a 
logical calculus, a formal mathematical theory, in which reasoning can be 
done based on formal logical symbol manipulation.

The resulting logical language can be applied to in depth, intrinsic analysis 
of the PF information. To begin with, such an analysis could be used to decide 
whether the relationships encoded in the firing patterns actually permit the
``spatial order'' interpretation. Specifically, the consistency of the 
{\em hypothesis} that the firing patterns of a certain collection of neurons 
(e.g. hippocampal or visual cortex cells) encode spatial relationships, in which 
cofiring of two cells represents the overlapping of the corresponding regions, 
can be conclusively tested using the RCC analysis in finite time, that grows 
polynomially with the number of the given ``cofiring''relationships. Such an 
approach provides an interesting example of a mathematical (logical) identification 
of the computational nature of the receptive fields. 

There exists a number of phenomenological models that aim to explain the
observed properties of the PF regions based on common qualitative relations
between PFs and the geometrical features of the environment (the ``geometric
determinants'') \cite{Burgess2,Burgess4,Hartley,OKeefe2,Touretzky3}. These
approaches exploit the experimentally observed trends in rat's spatial
behavior and its ``spatial instincts'', such as marking the behaviorally 
important areas (e.g. dead ends of tracks, food reward locations, etc.) as 
``places'', and use some phenomenological knowledge about the global behavior 
of the PFs \cite{Gothard2,Fenton,Muller1,Muller2,Muller3}. However, mere 
description of such PF's associations with corners and walls in itself does 
not provide a key for understanding the scope and the nature of the information 
contained in the hippocampal computations, as well as the reasoning structure 
based on it. However, the RCC topological calculus the analysis further to be 
taken further: given a certain initial amount of information, such as e.g., the 
association of several PFs with some ``geometric determinants'', the (I)SRE 
observer can reason logically about the sequence of PC activations in order to 
understand an animal's navigational strategies at the topological level, based 
on logical (arithmetical, mechanical) symbol manipulation, defined by the RCC
logical calculus.

\begin{figure}[tbp]
\begin{center}
\includegraphics{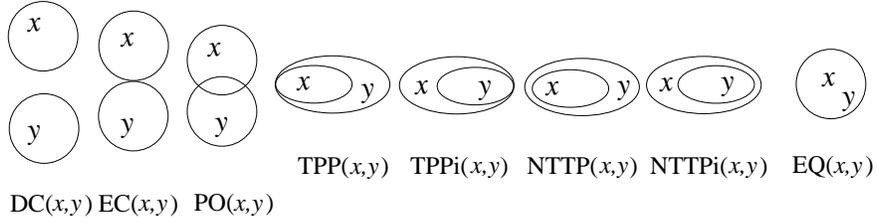}
\end{center}
\caption{RCC8 relationships.}
\label{rcc8}
\end{figure}

Within each particular RCC formalism, spatial order is defined via a family
of binary topological relations imposed on the regions, that guarantee the
consistency of the space constructed from them. The most widely used formulation 
of RCC\cite{Cohn1,Cohn2,Gotts,Randell} defines 8 jointly exhaustive and pairwise 
disjoint (JEPD) binary topological relations between pairs of regions with 
{\emph crisp} boundaries; this calculus is known as RCC8 and can be defined 
from the single primitive {\sf C}($x,y$) in a full first order theory, or the 
relations are taken to be primitive in a constraint language setting. These 
relations shown in Fig.\ref{rcc8} are: {\sf DC} ($x$ is disconnected from $y$), 
{\sf EC} ($x$ is externally connected with $y$), {\sf PO} ($x$ partially overlaps 
with $y$), {\sf TPP} ($x$ is a tangential proper part of $y$), {\sf NTPP} ($x$ is 
a nontangential proper part of $y$), {\sf TPPi} (inverse of {\sf TPP}), {\sf NTPPi} 
(inverse of {\sf NTPP}) and {\sf EQ}.
Axioms can be introduced to define regions corresponding to the
complement of a region, the sum, the difference and the product of a
pair of regions -- in the cases when they can {\em logically} exist.

Interestingly, these 
RCC relationships are also manifested at a cognitive level \cite{Knauff,Renz}.
In each PF layout, every pair of PFs can be related to one another via one of 
the JEPD relationships from a chosen RCC calculus, based on the analysis of the 
PC firing rates. The regions inferred via RCC Boolean functions may or may not 
be realized as actual PFs, however {\em logically} they are available for spatial 
reasoning both to the (I)SRE observer and to the animal, and hence they help the 
observer to follow spatial aspects of the animal's behavior and its navigational 
strategies. Hence RCC provides a practical tool for reconstructing the specific 
relationships between the regions coded in the hippocampal network and a clear 
logical scheme for spatial reasoning.
\begin{figure}[tbp]
\begin{center}
\includegraphics{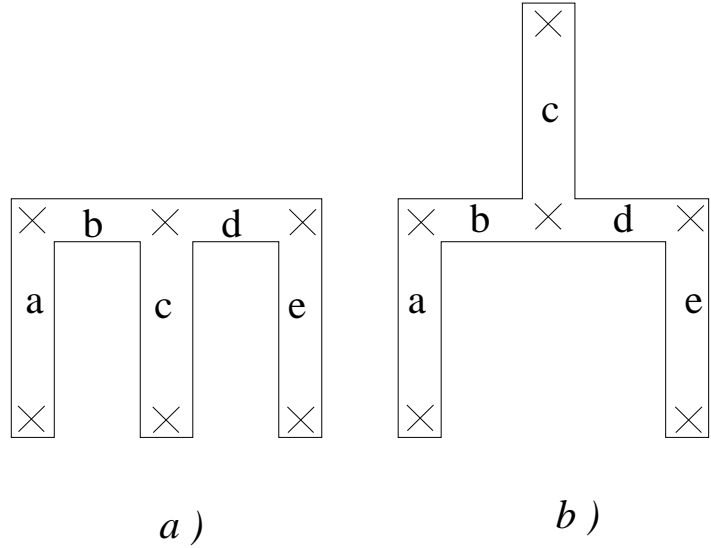}
\end{center}
\caption{W track, with 5 basic connected component parts and 6 marked points
(geometric determinants).}
\label{w_track}
\end{figure}
For example, the RCC8 based analysis can be applied to describing the 
spatial structure of the ``W'' linear track. As mentioned in Section 
\ref{section:partitions}, one of the biologically relevant distinguishing 
features of linear tracks is that some pairs of parts can be shrunk to a 
point whereas others cannot. For the 5 connected basic regions of the W 
track shown on Fig.\ref{w_track} a) that cannot be divided into two disconnected 
({\sf DC}) parts, the pairs $(a,b)$ or $(b,d)$ are contractible but $(a,c)$ or 
$(b,e)$ are not. The RCC8 relationships between the regions that describe the 
structure  of the W track environment are {\sf EC}($a$,$b$), {\sf EC}($b$,$c$), 
{\sf EC}($b$,$d$), {\sf EC}($c$,$d$), {\sf EC}($d$,$e$), {\sf DC}($a$,$c$), 
{\sf DC}($a$,$d$), {\sf DC}($a$,$e$), {\sf DC}($b$,$e$), {\sf DC}($c$,$e$).
Hence, the (I)SRE observer will immediately be able to infer the W 
(Fig. \ref{w_track} a) or topologically similar (Fig. \ref{w_track} b) structure 
of the track using RCC5 calculus, if the relationships 
between different pairs of (connected) PFs regions detected in the experiment (notably 
{\sf PO}($PF_{i}$,$PF_{j}$) and {\sf DR}($PF_{i}$,$PF_{j}$)) are compatible with these 
relationships. Given a sufficient number of PFs, it is also possible to determine the
structure of the environment in cases when the regions are multiply connected, however 
this case involves a more complicated analysis.

It is also conceptually important, that in the case of the PF, the regions 
``emerge'' from the discrete spiking patterns. Since at any moment of time $t$ 
and for any chosen frequency thresholds $\theta_{i}$ there exists only a finite 
number of spikes that define a region, the regions and their boundaries are 
``soft'', in the sense that the tangency relationships, such as ({\sf TPP} 
versus {\sf NTPP}), generally cannot be distinguished. As the statistical 
information about the PC firing accumulates with time, the regions associated with
PFs become sharper in static environments both to the observer and to the rat.
\begin{figure}[tbp]
\begin{center}
\includegraphics{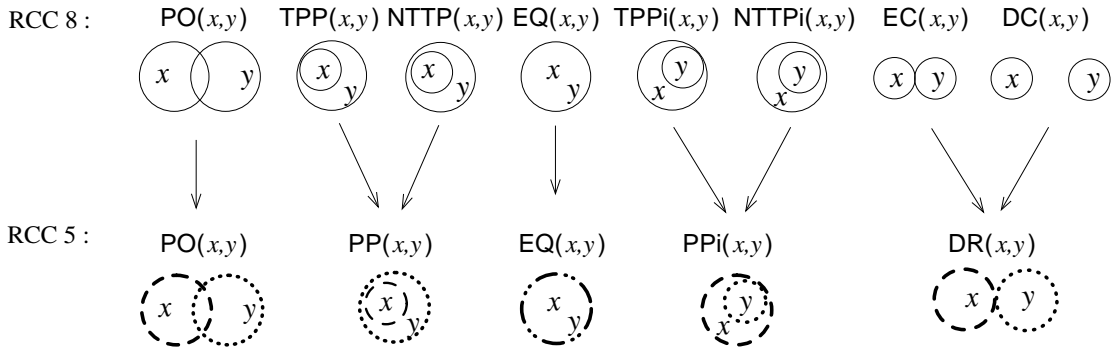}
\end{center}
\caption{RCC8 goes into RCC5. Dashed and dotted lines represent
region boundaries in RCC5 approach.}
\label{rcc_reduction}
\end{figure}
This argument suggests the of use of the RCC5 calculus, which is the result of ignoring 
tangency relationships in RCC8 and thus has a smaller set of JEPD relations to describe 
the relationship between two regions: {\sf DR} (discrete), {\sf EQ} (identical), 
{\sf PP} (proper part), {\sf PPi} (inverse {\sf PP}), and {\sf PO} (partial overlap), 
could be used instead -- see figure \ref{rcc_reduction}. Different choices of the 
threshold will generate a stack of the soft boundary regions that are related to 
one another via one of the RCC5 relationships.

It is important that the JEPD relationships between naturally ``fuzzy'' PF regions 
are also vague, in the sense that they emerge together with the PF themselves, as 
the number of spikes fired in a specific configuration accumulate. However, since
the data accumulation happens gradually, the logical study of the emergence (of
``crisping'', \cite{Cohn3}) of the regions {\em and} of the relationships between 
them can be made based on the analysis of the possible gradual transitions between 
the relationships. As shown in \cite{Cohn1,Cohn2,Randell}, the RCC relationships 
can be organized into natural succession sequences (conceptual neighborhoods) that 
specify the logical order in which RCC relationships between the regions can change.
This provides a possibility to follow the intriguing process of the emergence of
the region-defined space. 

Remarkably, RCC conceptual neighborhood analysis has the potential not only to
follow the development of the spatial relationships in static environments, 
but also to follow the changes in the PF configurations in the regime of 
smooth continuous transitions. Hence the (I)SRE observer can reason and resolve 
the space reconstruction task not only in static, but also in ``flexible'' tracks 
and arenas. 
\begin{figure}[tbp]
\begin{center}
\includegraphics{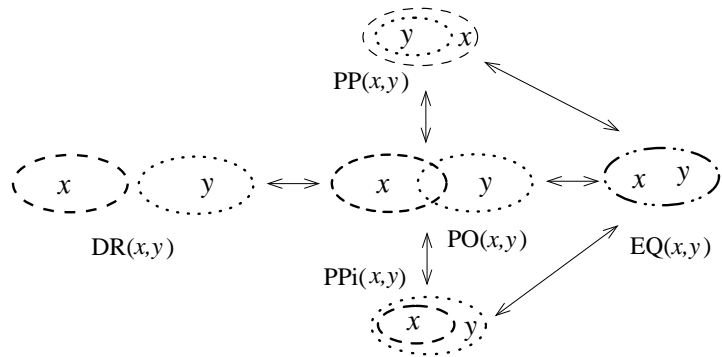}
\end{center}
\caption{The immediate conceptual neighborhood structure of RCC5 -- the possible 
sequences of gradual transformation of the binary JEPD relationships.}
\label{conceptual}
\end{figure}

In addition to ``natural'' fuzziness of the PF regions that is due to the
discreteness and stochasticity of spiking information, PFs are arbitrary defined 
by the randomly selected value of the firing threshold $\theta_{i}$. As mentioned 
above, any set of $n$ thresholds $f_{i,\alpha}\in \left[\theta_{i}^{\min},
\theta_{i}^{\max}\right]$ $\alpha=1$, ...,$n$, will produce a stack of $n$ soft 
boundary regions related to one another in RCC5 calculus via {\sf PP} or {\sf PPi} 
relationships. This ``stack of regions'' (even in the simplest case $n=2$) actually 
creates additional possibilities for the analysis within the so called ``egg-yolk 
theory'' \cite{Cohn3,Cohn96KRvague,Cohn94g,Lehmann94a} and its generalizations, 
which describes vague regions using a formal ``nested'' structure -- an inner 
``yolk'' representing space that is definitely part of the region (for the yolk
the threshold can be set sufficiently high) and the ``white'' between the yolk 
and the rest of the environment, representing space that may or may not be part 
of the PF. The threshold for the white (not-PFs) boundary can be set rather low 
to admit space that is not a part of the PF. Different egg-yolk versions of RCC5 
with varying numbers of JEPD relationships \cite{Cohn3} can be used for a specific 
logical analysis of a given PC data set.

\section{Discussion}
\label{section:discussion}

The analysis of the spatial tuning of the hippocampal PCs and the behavioral
manifestations of the hippocampal memory structure suggests that the 
representation of the information in the hippocampal system is based on explicit 
quasi-topological maps.
Overall, the existing experimental evidence seems to indicate that the hippocampus
encodes order relationships, which, in spatial context, translates directly into 
the topological order of the space in which the rat navigates. Although the PFs 
are tied to various geometrical features in the environment, e.g. corners, walls, 
objects on the arena, etc. \cite{OKeefe1}, the spatial information stored in the 
hippocampal network appears to be of a predominantly topological nature. 


The hippocampus is believed to perform indirect associations between various
memory and sensory patterns \cite{McNaughton4,Morris}, which leads to the 
appearance of a structural representation of the environment. In particular,
it appears that the animal internalizes a flexible discretized representation 
of space by ``memorizing'' an allocentric plastic grid of place fields. According 
to the view proposed above, the pointfree topology framework allows hippocampal 
activity in a stable regime to be interpreted as a manifestation of an emergent
quasitopological inner space, whereas the RCC calculus allows the use of the
information provided by the PCs to reason about its internal organization.
The power of the RCC method is that it provides a nontrivial logical
spatial reasoning scheme based on only a small number of PFs recorded in the
experiment, which makes it specially valuable for the real data analysis.

These principles of the spatial information representation and processing
reflect the general structure of memory organization. From the biological
point of view, the existence of a topological reference map stored in the
brain enables the animal to track the ``small changes'' of the sensory input
by putting them into a continuously defined context, as opposed to having to
evaluate anew every sensory configuration without recognizing patterns of
continuous change. Biologically, this may be related to the fact that an 
animal's survival depends on its ability to single out salient changes in
the environment as reliably and at the same time as flexibly as possible.
The animal must be ready to change its perception of the environment on the
time scale at which the external ``threats'' may occur. Often it is impossible
for the animal to know when and how its navigational task may change, hence
its behavioral decisions must be made based on spatial encodings which do
not precisely determine spatial location mechanisms \cite
{Bittner1,Bittner2,Cohn3,Hazarika}. It is nevertheless clear that having a
perfectly correct but vague solution to a fuzzily posed task is certainly
biologically more effective than spending time on producing computationally
costly and certainly inaccurate precise answer.

Generally, in order to understand the neural mechanism of space perception,
one needs to understand how the representation is created, learned,
adjusted, etc., as well as its functioning principles, i.e. how the
resulting map is used. The first question is addressed by various dynamical
or statistical network models that aim to reproduce, based on physiological
data, the overall functioning properties of the hippocampal network and its
interaction with the other brain parts \cite{Samsonovich2,Tsodyks2,Kali}.

Our analysis of the second question suggests a topological interpretation of 
the hippocampal spatial maps and their further understanding in the framework 
of the pointless topology and the QSR (RCC) spatial analysis and reasoning 
methods.

We are grateful to Sen Cheng for numerous discussions and valuable comments.

The work was supported in part by the Sloan and Swartz foundation and by the
NIH grant number F32 NS054425-01.

\section{Appendix. Physiological properties of the place cells}
\label{section:physiology}

Let us briefly outline some general physiological information concerning the
PCs and properties of PFs based on the rat data. Physiologically, the
hippocampal formation has several parts (CA1, CA3, Dentate Gyrus, and
Subiculum) which are subnetworks of different architecture and functionality 
\cite{Amaral1,Ishizuka,Traub2}. Cells in all of these parts have PC properties, 
however their spatial tuning (PF sizes, quality, responsiveness to external 
changes and so on) are different. Below we will discuss the PFs and PCs based 
mainly on the properties of the CA1 cells, which are the most commonly studied 
PCs of the hippocampus.

\begin{enumerate}
\item \qquad As a result of exploratory learning of the environment, the
hippocampal network falls (through some rapid plasticity mechanism \cite
{Bliss}) into a particular {\em global state} $S^{(i)}$, characterized by 
a particular set of intercellular connections strengths, which yield a 
specific kind of activity pattern \cite{Muller3}. In case of a linear track, 
PFs appear after 4-5 mins (a few runs across a few meter long track) and in 
open arenas of similar size PF form in 10-25 minutes and take a few days to 
stabilize. The settling of the network into a particular state is often 
described by various attractor network \cite{Tsodyks1,Tsodyks2,Samsonovich1} 
or statistical (spin glass) models \cite{Fuhs1,Treves} in which each state 
$S^{(i)}$, $i=1$, ..., is represented by a local (in parameter space) dynamical 
regime or local minima of appropriate statistical weight functionals. In either 
case, the existence of (dynamical or statistical) basin of attractions references 
the fact that as long as the outside world provides a steady flow of incoming 
sensory stimuli that does not imply any significant changes of the environment 
for the rat, the hippocampus as a whole remains in the same regime in which 
every cell ``knows its place'' with respect to other cells in the network, 
and the sensory input simply triggers the firing events. The size of the 
basin of attraction (depth of the local minimum) and the stability of 
$S^{(i)}$ depend on many factors, e.g. on how long the animal was learning 
the environment, in what sequence \cite{Wills}, etc.

\item \qquad It is clear that different parts of the environment can be
distinguished from each other because each location is characterized by its
own combination of physical (or idiothetic) cues, which produce a particular
set of sensory inputs that the rat receives at (or in the course of getting
to) that location. However, it is important that the rat can map the same
environment in many different ways \cite{Fuhs3,Tanila}, so the same
combination of sensory inputs may produce entirely different patterns of
hippocampal activity in the same rat, depending on the state of its
hippocampal network. Moreover, the PCs can retain their firing regime even
if only a partial sensory input (compared to e.g. the sensory input when the
environment was being learned by the animal) is present. For example, the
animal can retain its system of PFs if the lights are dimmed or even in
complete darkness \cite{McNaughton1}. Hence, although in wake animals it is
primarily the sensory input that triggers the PC activity, however it is not
just a simple ``funneling'' of the sensory input into the hippocampus that is
responsible for the firing of the PCs. PCs can also fire in the right
navigational sequences during spontaneous network activations while the
animal is asleep \cite{Foster,Louie,Skaggs,Wilson1}. To emphasize this
point, below the state $S^{(i)}$ will be referring to the overall {\em regime} 
in which the network operates, its global configuration, as opposed to the
current spiking activity level of the hippocampal cells. Certainly, the
state of the network can also change, but this change happens on a different
time scale during a catastrophic restructuring (remapping) event.

\item \qquad The structure of $S^{(i)}$ imposes a particular discretization 
scheme onto the sensory input and a particular sequence of the firing activity 
shift from cell to cell. Hence the activity of the hippocampal network in the 
state $S^{(i)}$ produces a partition of the environment into a discrete collection 
of overlapping regions (PFs), so $S^{(i)}$ encodes the environment in an explicit 
{\em global}, i.e. simultaneously defined for the whole environment, map.

\item \qquad The sizes and the shapes of the PFs corresponding to different
PCs can vary significantly. The same cell in different environments or in a
different space mapping state $S^{(i)}$ in the same environment
can produce PFs of completely different sizes and shapes. Typically linear
sizes of PFs range between $10$ to about $75$ $cm$ \cite{Best1}. In certain
cases PFs can be as big as the size of the whole environment accessible to
the animal -- e.g. $10$ meters in diameter. Typically PFs are compact, convex
(convexity and other shape features of PFs can be modulated by the
environment, e.g. by a curving of a track) regions of space. In some cases
PCs can have multiply connected PFs, however one can argue that in an
environment perceived by the animal as a single integral piece, each PC has
a single, compact, contractible PF. No two PFs of different cells entirely 
coincide or even have same maxima locations, so each cell has a unique place
field inside the environment. In linear environments, a PF can be directional, 
i.e. a PC may be active only if the rat moves in one of the two possible 
directions, but not in the other.

\item \qquad The peak firing frequency (about $20$ Hz) of a particular cell 
$c$ is reached when a rat passes through the center of its $PF_{c}$, however
peak firing rates may differ from cell to cell and for a given cell from
environment to environment. Outside of its PF the cell shows only some weak
``background noise activity'' (about $0.1$
Hz). The averaged spatial firing frequency (the firing probability) of a PC
as a function of the coordinates $x$ and $y$ produces a surface that often
looks like and usually is modeled as a Gaussian hump over a compact base.
As long as the rat remains in the same location, the activity level of the
PC remains approximately constant \cite{Best1}.

\item \qquad A given point in space may be shared by many PFs. At a given 
location, about $300$ out of $3\cdot 10^{5}$ of CA3 cells and $3000-4000$ 
out of $4\cdot 10^{5}$ of CA1 cells are simultaneously active, however the 
activity of only few of them are near maximal level 
\cite{Amaral2,Ishizuka,Samsonovich1}.

\item \qquad 
\label{allocentric}
One of the most important properties of the PFs is that they are 
{\em allocentric,} i.e. the spatial location of the PFs is fixed in the external 
world coordinate frame, regardless of the history of animal's behavior. After 
the animal familiarizes itself with the (static) environment and creates its 
PF partition, the location of each PF does not change significantly between 
the rat's visits. In a given state of the network, it is always the same cell 
that marks a given place field.

Overall, the PFs' patterns are not significantly influenced by the details 
of animal's behavior, the PFs layout structure can be basically ``decoupled''
from the animal's behavior, and hence it can be used for the analysis of the
environment, rather than a combination of environmental features and the internal
state of animal's brain.

\item \qquad The external location of the PF of a given cell in different maps
can be very different. It is not fixed either with respect to other PFs or with 
respect to any particular ``features'' of the environment. Generally speaking, 
if the animal is moved from one environment $E_{1}$ into another $E_{2}$, then 
the PF partition of $E_{2}$ will be entirely different from that of $E_{1}$, 
even if both environments have similar features \cite{Sharp}. The old ``corner'' 
cells may mark the center of the new arena or appear in some other corner or 
stick along some wall segment. It seems impossible to predict from previous 
behavior where or even whether a cell will have its PF in the new environment 
before it actually appears. For a given cell in a given environment, there is 
about 30\% chance of being active \cite{Samsonovich1,McNaughton3}.

\item \qquad Generally speaking, there is no commonality between the spatial
position of PFs in different rats who have learned the same maze. Every rat
comes up with its own mapping, i.e. its own assignment of ``places'' to cells
(or visa versa). Overall it appears that finding a place for the PF in a given 
environment is a very flexible dynamical process with many parameters.

\item \qquad
\label{partition}
The pattern of the positions of PCs in the brain tissue (i.e. in the CA1 or the 
CA3 areas) does not correspond to the spatial pattern of the PFs in the environment. 
In particular, neighboring cells do not in general map to neighboring physical 
places. For example, the PFs of a group of nearby cells recorded from a single 
electrode are scattered all over the environment. In this sense, the PF mapping 
of the environment to the cells of the hippocampus is not ``topographical''. 
If parts of two place fields $PF_{c_{1}}$ and  $PF_{c_{2}}$ overlap, it does not 
mean that one cell can ``activate'' the other. If the rat enters $PF_{c_{1}}$ 
and has not yet entered $PF_{c_{2}}$, with $PF_{c_{1}}\cap PF_{c_{2}}\neq \varnothing$, 
then the cell $c_{2}$ will be silent until the rat actually gets inside $PF_{c_{2}}$.
As the number of the implanted
electrodes and hence the recorded cells increases (currently about 150 cells
can be recorded simultaneously \cite{Wilson2}), the corresponding PFs
produce progressively denser cover of the environment. The union of all PFs
completely covers the environment $E$,
\begin{equation}
\cup_{c}PF_{c}=E.
\end{equation}
However, there is a difference in PF occupancy for ``interesting'' places
compared to ``uninteresting'': there are typically more PFs around food wells,
maze junctions and other ``important places''.

\item \qquad An important ensemble characteristics of PC firing activity are 
the global oscillations of the intracellular potential, in particular the 8 Hz 
$\theta$-oscillations \cite{Buzsaki1,Buzsaki2,Traub1}, as well as the correlations 
between the local phase of $\theta$-rhythm and the firing rate of a PC \cite{OKeefe5}. 
These oscillations play an important role in synchronization across large populations of 
cells, and often help to understand the functional connections between the activity 
patterns of different cells \cite{Hasselmo1,Hasselmo2,Lisman,Lisman1}.

\end{enumerate}

Based on these properties of the PCs and on some additional observations of
the PF ensemble behavior described below, the goal of the following
discussion will be to provide some arguments in favor of the topological
nature of the spatial representations generated by hippocampal neural
activity.

\end{document}